\documentclass[preprint,aps,showpacs,superscriptaddress,preprintnumbers,amsmath,amssymb]{revtex4}

\usepackage{graphicx}
\usepackage{dcolumn}
\usepackage{bm}

\begin{document}

\preprint{APS/123-QED}

\title{Instability of ion kinetic waves in a weakly ionized plasma}

\author{Roman Kompaneets}
\affiliation{School of Physics,
The University of Sydney, New South Wales 2006, Australia}

\author{Alexei V. Ivlev}
\affiliation{Max-Planck-Institut f\"ur extraterrestrische Physik, 85741 Garching, Germany}

\author{Sergey V. Vladimirov}
\affiliation{School of Physics, The University of Sydney, New South Wales 2006, Australia}

\author{Gregor E. Morfill}
\affiliation{Max-Planck-Institut f\"ur extraterrestrische Physik, 85741 Garching, Germany}


\begin{abstract}
The fundamental higher-order Landau plasma modes are known to be generally heavily damped. 
We show that these modes for the ion component in a weakly ionized plasma can be 
substantially modified by ion-neutral collisions and a dc electric field driving ion flow
so that some of them can become unstable.
This instability is expected to naturally occur in 
presheaths of gas discharges at sufficiently small pressures and thus affect sheaths and discharge structures.
\end{abstract}

\pacs{52.35.Fp, 52.25.Ya, 52.30.-q, 52.25.Dg}
\maketitle

\section{Introduction}
One of the fundamental kinetic phenomena in plasma physics is the higher-order Landau modes~\cite{derfler-sim-phys.fluids-1969}.
They are the heavily damped solutions of the
dispersion relation describing the electrostatic modes of a one-component collisionless
Maxwellian plasma~\cite{landau-j.phys.ussr-1946}.
(A one-component plasma is the approximation where only one plasma component
oscillates.)
The dispersion relation is:
\begin{equation}
1+\frac{1}{(k \lambda_{\rm D})^2}\left[ 1+\frac{\omega}{k v_{\rm t} \sqrt{2}} Z \left( \frac{\omega}{k v_{\rm t}\sqrt{2}}\right)\right]=0,
\label{landau-dispersion-relation}
\end{equation}
where 
\begin{eqnarray}
Z(x)= \frac{1}{\sqrt{\pi}} \int \limits_U \frac{\exp(-\xi^2) \, d\xi}{\xi-x} \nonumber \\
\equiv 2i \exp(-x^2)\int_{-\infty}^{ix} \exp(-\xi^2) \, d\xi
\label{plasma-dispersion-function}
\end{eqnarray}
is the plasma dispersion function, 
$U$ is any contour in the complex $\xi$ plane
passing from $\xi = -\infty$ to $+\infty$ below the singular point $\xi=x$,
$\omega$ is the complex wave frequency, $k$ is the real wave number,
$v_{\rm t}=\sqrt{k_{\rm B}T/\mu}$ is the thermal velocity,
$T$ is the temperature, $\mu$ is the particle mass, $k_{\rm B}$ is the Boltzmann constant,
$\lambda_{\rm D}=v_{\rm t}/\omega_{\rm p}$ is the Debye length,
and $\omega_{\rm p}$ is the plasma frequency.
The transcendental equation (\ref{landau-dispersion-relation})
with respect to $\omega$ yields the Langmuir mode and an infinite number of the heavily damped higher modes~\cite{derfler-sim-phys.fluids-1969}. 

Because the higher modes are a fundamental phenomenon, it is not surprising that
they received considerable attention in the literature despite their strong damping. 
The first, but implicit, mention dates back to Landau himself \cite{landau-j.phys.ussr-1946}, who used terms ``all the poles'' and ``that of the poles'' in relation to the solutions of 
Eq.~(\ref{landau-dispersion-relation}). 
An explicit statement on the existence of the higher modes was made 14 years later by Jackson, who demonstrated their presence analytically \cite{jackson-j.nucl.energy-1960}. 
Numerical results were published in the 1960s \cite{fried-gou-phys.fluids-1961, denavit-phys.fluids-1965, derfler-sim-phys.fluids-1969}. 
Recently, these modes have been studied in relativistic plasmas \cite{kneller-sch-j.plasma.phys-1998, shcherbakov-phys.plasmas-2009}. 
Experimentally, the higher modes were observed in the 1970s (in the spatially damped case) \cite{kawai-ike-har-plasma.phys-1976,
ikezawa-kaw-har-j.plasma.phys-1977}.

Very similar, but nonlinear, kinetic electron waves have been recently observed in simulations to persist 
without any apparent decay over many plasma periods after being excited by an artificial external driver
\cite{afeyan, johnston-tys-ghi-phys.plasmas-2009}. They were called KEEN waves,
standing for ``kinetic electrostatic electron nonlinear'' waves~\cite{afeyan, johnston-tys-ghi-phys.plasmas-2009}.
One remarkable similarity between them and the higher modes is that
KEEN waves were obtained even for frequencies well below the plasma frequency and that their typical phase velocity
was somewhat above the electron thermal velocity,
which is in full accord with the properties of the higher modes.
Another similarity can be seen in the quasineutral character of KEEN waves
(evidenced by Fig.~9 of Ref.~\cite{johnston-tys-ghi-phys.plasmas-2009}),
as the higher modes are quasineutral at small wave numbers
[i.e., the neglect of the vacuum term in Eq.~(\ref{landau-dispersion-relation})
does not affect the higher solutions $\omega$ at small wave numbers,
as can be easily verified].
The heavy Landau damping, characteristic to the higher modes, 
was eliminated for KEEN waves by a population of trapped particles created by the external drive. 
Very recently, analogous nonlinear kinetic waves have been studied for the ion component in the presence of the electron response
using a similar external driver \cite{valentini-cal-per-phys.rev.lett-2011}.

In this paper, we show that a finite number of the higher modes can become unstable 
under a quite natural set of circumstances, so that kinetic waves can 
emerge without excitation by an artificial external driver.
Namely, we show that the higher modes for the ion component in a weakly ionized plasma can be 
substantially modified by ion-neutral collisions and ion flow driven by a dc electric field
so that some of these modes can become unstable.
Such flows are an essential feature of low-pressure gas discharges~\cite{zeuner-mei-vacuum-1995,
hershkowitz-ko-wan-ieee.trans.plasma.sci-2005, oksuz-her-plasma-plasma.sources.sci.tech-2005,
land-goe-new.j.phys-2006, nosenko-fis-mer-phys.plasmas-2007}, as they naturally arise to maintain the balance of absorption of ions and 
absorption of electrons
on the walls and electrodes of the discharge chamber.

So far, there have been numerous investigations of streaming instabilities
triggered due to relative flows of various plasma components in collisionless plasmas,
with perhaps the most known example being the Buneman instability \cite{buneman-phys.rev-1959}, but our study is principally different
in two important aspects explained below.

The first aspect is that we include a dc electric field and collisions with neutrals, and do this {\it self-consistently}. 
The self-consistency here means two things. First, we find the steady-state velocity distribution from the model itself, i.e., from 
the balance of ion acceleration in the field and ion-neutral collisions (instead of assuming a model distribution, e.g., a shifted Maxwellian distribution).
Second, collisions and the dc field
not only define the steady state but are also fully accounted for in the analysis of perturbations
(this is essential to obtain the instability, as we show in Sec.~\ref{interpretation}).
Of course, such an approach generally requires extremely cumbersome velocity calculations,
but we avoid this difficulty by considering the common case where the dominant mechanism of collisions is charge transfer, in which
the ion and neutral simply exchange identities and thus velocities~\cite{raizer-1991, lieberman-1994}. Our further approximation is to assume
the collision frequency to be velocity independent, which allows an elegant solution not only
for the steady state but also for the ion susceptibility \cite{ivlev-zhd-khr-phys.rev.e-2005, kompaneets-derivation}. As, in reality, 
it is not the collision frequency but the cross section that 
is characterized by a weak (logarithmic) velocity 
dependence \cite{lieberman-1994, smirnov-2001-p75} (in the regime where charge transfer is the dominant mechanism of collisions), 
we separately show that the instability
remains in the constant mean free path case (in Sec.~\ref{role-of-the-form-of-the-collision-term}). 

The second aspect is that we consider {\it only one} oscillating plasma species, i.e., ions.
The electron density is assumed to be fixed, which 
physically implies that the temperature of the electrons is high enough so that they do not ``feel'' electric fields, similar to how ion Langmuir waves are derived in 
classical textbooks \cite{lifshitz-pitaevskii-1981}.
This assumption allows us to render a simple physical picture,
while we address the role of the electron temperature in Secs.~\ref{electron-response} and
\ref{model-applicability-limits} (with additional details given in Appendix \ref{estimates})
and provide explicit conditions for neglecting the electron 
effects [Eqs. (\ref{limitation1}), (\ref{instability-restriction}), (\ref{range-electron-response}), and (\ref{compatibility-1})]. 
As our model involves only one oscillating plasma component, it is remarkable indeed
that our analysis reveals an instability. This instability is clearly associated with a novel mechanism.

We emphasize that the ion-kinetic instability described in this paper is not a variation of 
any known collisionless instability (e.g., the bump-on-tail instability \cite{melrose-1986, nishikawa-wakatani-2000-p.92}),
since a collisionless one-component plasma with our steady-state velocity distribution is always stable (Sec.~\ref{interpretation}). 
We explain the instability mechanism in Sec.~\ref{interpretation}.

This ion-kinetic instability should affect a large class of gas discharges, as it
is expected to naturally occur in presheaths \cite{riemann-j.phys.d.appl.phys-1991, riemann-j.phys.d.appl.phys-2003,
hershkowitz-ko-wan-ieee.trans.plasma.sci-2005, 
oksuz-her-plasma-plasma.sources.sci.tech-2005, 
hershkowitz-phys.plasmas-2005} under quite common conditions (Sec.~\ref{presheaths}).
In even broader context, our study shows that the often disregarded higher-order 
Landau modes can in fact play a crucial role in the presence of an electric field.

\section{Methods}

\subsection{Basic equations}
Let us consider a weakly ionized plasma in a dc electric field ${\bf E}_0$ driving ion flow. 
For electrons we assume a Boltzmann distribution with a sufficiently large temperature
so that we can consider their number density $n_0$ to be homogeneous and fixed
(note that Boltzmann electron distributions in the presence of field-driven ion flow
are common in discharges \cite{riemann-j.phys.d.appl.phys-1991, riemann-j.phys.d.appl.phys-2003,
hershkowitz-ko-wan-ieee.trans.plasma.sci-2005,
oksuz-her-plasma-plasma.sources.sci.tech-2005,
hershkowitz-phys.plasmas-2005}; we address the role of the electron temperature
in Secs.~\ref{electron-response} and
\ref{model-applicability-limits}).
We assume ${\bf E}_0$ to be homogeneous and use
the kinetic equation for ions 
with the Bhatnagar-Gross-Krook (BGK) ion-neutral collision term \cite{ivlev-zhd-khr-phys.rev.e-2005} and Poisson's equation:
\begin{eqnarray}
\frac{\partial f}{\partial t}
+{\bf v} \cdot \frac{\partial f}{\partial {\bf r}}
+\frac{e}{m} \left( {\bf E}_0 - \frac{\partial \phi}{\partial {\bf r}} \right)
\cdot \frac{\partial f}{\partial {\bf v}} \nonumber \\
 = -\nu f +\nu \Phi_{\rm M} \int f({\bf v}') \, d{\bf v}',
\label{kinetic-equation}
\end{eqnarray}
\begin{equation}
-\bigtriangleup \phi= \frac{e}{\epsilon_0} \left( \int f \, d{\bf v} -n_0 \right),
\label{poisson}
\end{equation}
where $f$ is the ion distribution function,
$\phi$ is the electric potential
describing the time-space varying field
(i.e., the field apart from ${\bf E}_0$),
\begin{equation}
\Phi_{\rm M}({\bf v})=\frac{1}{(2\pi v_{\rm tn}^2)^{3/2}}\exp\left(-\frac{v^2}{2v_{\rm tn}^2}\right)
\label{neutral-distribution}
\end{equation}
is the normalized Maxwellian velocity distribution of neutrals,
$\nu$ is the velocity-independent ion-neutral collision frequency,
$v_{\rm tn}=\sqrt{k_{\rm B}T_{\rm n}/m}$ is the thermal velocity of neutrals,
$T_{\rm n}$ is the temperature of neutrals,
$e$ is the elementary charge (ions are assumed to be singly ionized),
$m$ is the ion mass, and $\epsilon_0$ is the permittivity of free space.
The BGK term exactly 
describes charge transfer collisions under the assumption
of a velocity-independent collision frequency \cite{ivlev-zhd-khr-phys.rev.e-2005}.

\subsection{Steady state}

\begin{figure}
\includegraphics[width=10cm]{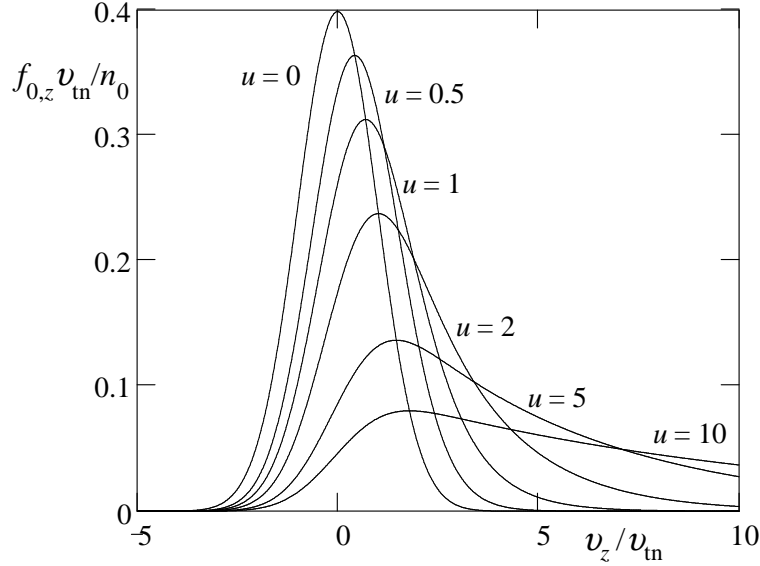}
\caption{Steady-state solution. Shown is the longitudinal velocity distribution
$f_{0,z}=\int_{-\infty}^{\infty} \int_{-\infty}^{\infty} f_0 \, d v_x \, dv_y$,
where the $z$ axis is in the direction of the field.
Different lines
correspond to different values of the 
parameter $u={v_{\rm f}}/v_{\rm tn}$.}
\label{fig-distribution}
\end{figure}

The homogeneous steady-state solution $f=f_0$ is found from
Eqs.~(\ref{kinetic-equation}) and (\ref{poisson}) by setting $\phi=0$, $\partial f/\partial t =0$, $\partial f/\partial {\bf r}={\bf 0}$.
This gives~\cite{sugawara-tag-sak-j.phys.d.appl.phys-1996, ivlev-zhd-khr-phys.rev.e-2005}:
\begin{equation}
f_0({\bf v})= \frac{n_0} {(2\pi v_{\rm tn}^2)^{3/2}}\int_0^{\infty}\exp\left(-\xi -\frac{|{\bf v}-\xi{\bf v}_{\rm f}|^2}{2v_{\rm tn}^2}\right) \, d{\xi},
\label{distribution}
\end{equation}
where 
\begin{equation}
{\bf v}_{\rm f}= \frac{e{\bf E}_0}{m\nu}
\end{equation}
and the subscript ``f'' stands for ``flow,'' as the flow velocity $(1/n_0)\int {\bf v} f_0 \, d{\bf v}$ can be shown to be equal to
${\bf v}_{\rm f}$.
It is helpful to note that Eq.~(\ref{distribution}) here is simply Eq. (3) of
Ref.~\cite{ivlev-zhd-khr-phys.rev.e-2005}, but rewritten using another integration variable
in order to show that $f_0$ is an integral superposition of shifted Maxwellian distributions with exponential
weights. At $E_0=0$, the velocity distribution (\ref{distribution}) is Maxwellian, with 
the thermal velocity being equal to that of neutrals.
The solution (\ref{distribution}) is shown in Fig.~\ref{fig-distribution}.
As Fig.~\ref{fig-distribution} shows, at large fields the solution~(\ref{fig-distribution}) cannot be approximated by a shifted Maxwellian distribution, 
as the solution~(\ref{distribution}) becomes highly asymmetric with respect
to the position of its maximum.
This can also be illustrated by considering the limit of 
cold neutrals, $v_{\rm tn} \to 0$, where
Eq.~(\ref{distribution}) becomes
\begin{eqnarray}
f_0({\bf v})=\frac{n_0}{v_{\rm f}} \exp \left( -\frac{v_z}{v_{\rm f}} \right) \delta(v_x) \delta (v_y), \quad v_z>0, \nonumber \\
f_0 ({\bf v})=0, \quad v_z<0,
\label{simple-distr}
\end{eqnarray}
where the $z$ axis is in the direction of ${\bf E}_0$.

\subsection{Dispersion relation}

The dispersion relation is obtained by linearizing Eqs.~(\ref{kinetic-equation}) and (\ref{poisson})
with respect to $\phi$ and $f-f_0$ and can be written using the ion susceptibility derived in Refs.~\cite{ivlev-zhd-khr-phys.rev.e-2005, schweigert-plasma.phys.rep-2001}.
A derivation of this dispersion relation by solving the initial value problem is given in Ref.~\cite{kompaneets-derivation}.
The dispersion relation is:
\begin{subequations}
\label{dispersion-relation}
\begin{equation}
1+\frac{\omega_{\rm pi}^2}{\nu^2}
\frac{B(\omega, {\bf k})}{1-A(\omega, {\bf k})}=0, 
\label{dis-1} 
\end{equation}
\begin{equation}
A(\omega, {\bf k})=\int_0^\infty \exp [-\Psi(\omega, {\bf k}, \eta)] \,d{\eta}, \label{dis-2}
\end{equation}
\begin{equation}
B(\omega, {\bf k})=\int_0^\infty 
\frac{
\eta \exp [-\Psi(\omega, {\bf k}, \eta)]} 
{1+i({\bf k} \cdot {\bf v}_{\rm f}/\nu)\eta }
\,d{\eta}, \label{dis-3}
\end{equation}
\begin{eqnarray}
\Psi(\omega, {\bf k, \eta})=
\left(1-\frac{i\omega}{\nu}\right)\eta \nonumber \\
+\frac{1}{2}
\left[
\frac{i{\bf k} \cdot {\bf v}_{\rm f}}{\nu} 
+\left(\frac{kv_{\rm tn}}{\nu}\right)^2
\right]
\eta^2, \label{dis-4} 
\end{eqnarray}
\end{subequations}
where $\omega_{\rm pi}=\sqrt{n_0 e^2/(\epsilon_0 m)}$ is the ion plasma frequency, $\omega$ is the complex wave frequency,
and $k$ is the real wave number; the solutions $\omega$ of Eq.~(\ref{dispersion-relation}) provide contributions 
$\propto \exp(-i\omega t + i {\bf k} \cdot {\bf r})$
to the asymptotic expression for the solution $\phi = \phi({\bf r}, t)$ of the initial value problem at large $t$~\cite{kompaneets-derivation}.

The result (\ref{dispersion-relation}) is different from what one obtains by simply substituting 
our steady-state distribution~(\ref{distribution}) to the classical expression for the dielectric function of a collisionless plasma~\cite{lifshitz-pitaevskii-1981}.
The difference is due to our accounting for the perturbation term $(e{\bf E}_0/m)\cdot \partial (f-f_0)/\partial {\bf v}$
and the perturbation of the right-hand side of Eq.~(\ref{kinetic-equation}). It is this difference that results in the instability, as shown in Sec.~\ref{interpretation}.

\subsection{Analysis}
\label{analysis}

First, we numerically analyze the dispersion relation (\ref{dispersion-relation}) (in Sec.~\ref{numerical-results}).
The corresponding dimensionless variables are the {\it flow parameter}
\begin{equation}
u = \frac{v_{\rm f}}{v_{\rm tn}},
\label{flow-parameter}
\end{equation}
the {\it collision parameter}
\begin{equation}
\zeta=\frac{\nu}{\omega_{\rm pi}},
\end{equation}
the dimensionless frequency $\omega/\omega_{\rm pi}$ and
the dimensionless wave number ${\bf k} \lambda$, where 
\begin{equation}
\lambda=\frac{v_{\rm tn}}{\omega_{\rm pi}}.
\label{lambda}
\end{equation}
Note that in the above definition of $\lambda$ we use the thermal velocity of neutrals,
so at $E_0=0$ the length $\lambda$ becomes the ion Debye length.
For finite $E_0$, the ion thermal velocity and the ion Debye length are not defined
because of the non-Maxwellian form of the velocity distribution.
Also note that the length $\lambda$ is not the effective screening length in the presence of the flow,
as evident from Ref.~\cite{kompaneets-kon-ivl-phys.plasmas-2007}.
The dimensionless form of the dispersion relation (\ref{dispersion-relation}) is given in Appendix~\ref{dispersion-relation-dimensionless}. 

Second, we provide an analytical proof of the instability existence using Eq.~(\ref{dispersion-relation}) (in Sec.~\ref{proof})
and explain the instability mechanism (in Sec.~\ref{interpretation}).

Third, we analyze whether the instability remains 
in the constant mean free path case (in Sec.~\ref{role-of-the-form-of-the-collision-term}). For this purpose,
we replace 
the right-hand side of Eq.~(\ref{kinetic-equation}) by
\cite{else-kom-vla-phys.plasmas-2009}:
\begin{eqnarray}
{\rm St}[f({\bf r}, {\bf v})]
      = \int \frac{|{\bf v}' - {\bf v}|}{\ell}
      \left[\Phi_{\rm M}({\bf v}) f({\bf r}, {\bf v}') - \Phi_{\rm M}({\bf v}') f({\bf r}, {\bf v})\right]
      d{\bf v}',
\label{cmfp}
\end{eqnarray}
where $\ell$ is the collision length.  This operator exactly 
describes charge transfer collisions under the assumption
of a velocity-independent cross section.

Finally, we numerically study the effect of the electron response (in Sec.~\ref{electron-response}). 
To clarify, there are two effects related to a finite electron temperature: (i) the electron response to ion oscillations,
and (ii) a finite inhomogeneity length of the Boltzmann electron distribution in the field ${\bf E}_0$.
Here we focus on effect (i)
by adding
the Boltzmann response term
$1/(k \lambda_{\rm e})^2$
to the left-hand side of Eq.~(\ref{dis-1}), 
where $\lambda_{\rm e}=[\epsilon_0 k_{\rm B} T_{\rm e}/(n_0 e^2)]^{1/2}$ is the electron Debye length
and $T_{\rm e}$ is the electron temperature. 
Effect (ii) is discussed in Sec.~\ref{model-applicability-limits}.

\section{Results}
Let us briefly summarize our results before describing them in detail:

\begin{enumerate}

\item The instability occurs when $u \gtrsim 8$ and $\zeta \lesssim 0.3$ (Sec.~\ref{numerical-results}). However, in the limit $\zeta \to 0$, $u = const$ 
[physically corresponding to a collisionless one-component plasma with our steady-state velocity distribution (\ref{distribution})]
the instability growth rate tends to zero (Sec.~\ref{numerical-results}, Sec.~\ref{interpretation}).

\item It is downstream waves that become unstable at the above instability boundary (Sec.~\ref{numerical-results}). To clarify,
by ``downstream waves'' we mean that the phase velocity vector ${\rm Re}(\omega){\bf k}/k^2$
is in the direction of ${\bf E}_0$. 

\item When the above instability conditions are met, the instability occurs in a finite range of wave numbers.
Its lower and upper ends for downstream waves can be estimated 
as $k \sim \nu/v_{\rm f}$ and $k \sim (\nu/v_{\rm f}) {\rm min}\left\{u^2, \zeta^{-4/3}\right\}$, respectively
[Sec.~\ref{interpretation}].

\item The number of unstable modes depends on $u$ and $\zeta$
and can in principle be made arbitrarily high (Sec.~\ref{proof}).

\item The frequency of the most unstable mode for downstream waves is $\omega \approx (3.35 + 0.64i) \sqrt{eE_0 k/m}$, provided
that the above instability conditions on $u$, $\zeta$ and $k$ are satisfied by a considerable margin (Secs.~\ref{proof} and \ref{interpretation}).

\item The physics of the resulting growing waves is essentially kinetic (Sec.~\ref{interpretation}). 

\item The instability remains in the constant mean free path case (Sec.~\ref{role-of-the-form-of-the-collision-term}).

\item The electron response for typical values of $T_{\rm e}$
does not shift the above instability thresholds to unrealistic values,
as for $T_{\rm e}/T_{\rm n} = 200$ (corresponding to $k_{\rm B}T_{\rm e} = 5$~eV,
$T_{\rm n}=300$~K) the first higher mode remains unstable at, for instance, $u=14$ and $\zeta=0.05$ (Sec.~\ref{electron-response}).

\end{enumerate}
Let us now describe these findings in detail.

\subsection{Numerical results}
\label{numerical-results}
This section provides the results of the numerical analysis of the dispersion relation~(\ref{dispersion-relation}).

\begin{figure}
\includegraphics[width=11cm]{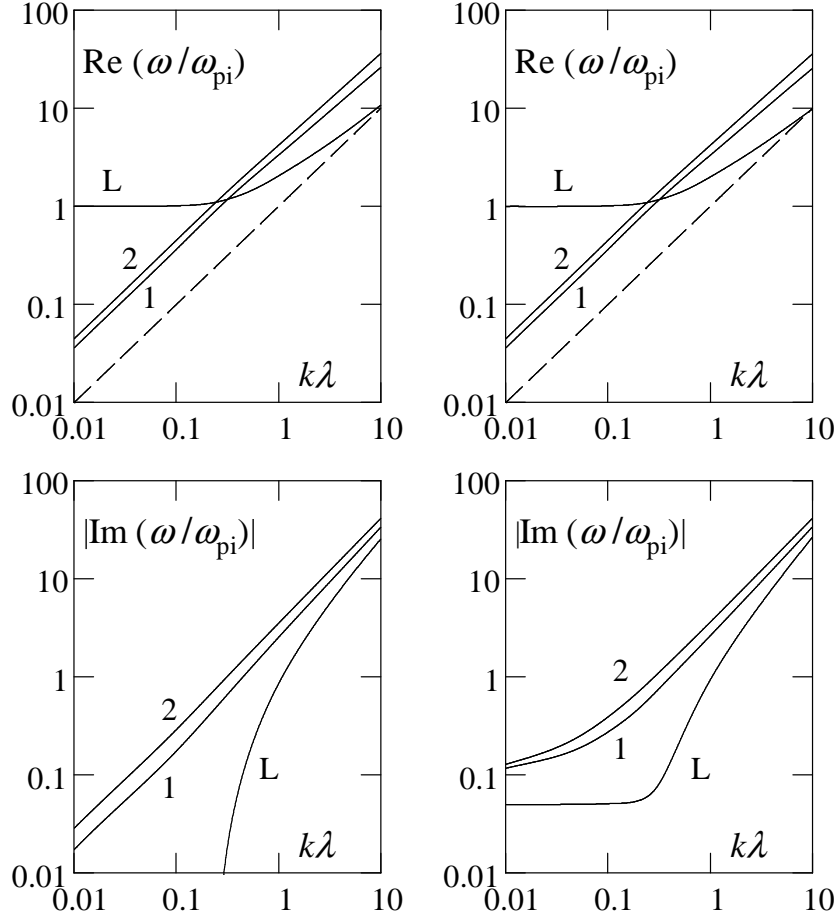}
\caption{Modes in the absence of flow.
Shown are the solutions of Eq.~(\ref{dispersion-relation}) for $u=0$.
The left column represents the collisionless case [$\zeta \to 0$; in this case
the dispersion relation is reduced to Eq.~(\ref{landau-dispersion-relation})], 
and the right column
illustrates the effect of collisions for $\zeta=0.1$.
The ion Langmuir mode and the first two higher modes are 
denoted by ``L,'' ``1,'' and ``2,'' respectively. The dashed lines, shown to guide the eye, 
correspond to ${\rm Re}(\omega)/k=v_{\rm tn}$.}
\label{u0}
\end{figure}

{\it No-flow case:} At $u=0$ and $\zeta \to 0$,
Eq.~(\ref{dispersion-relation}) is equivalent
to the Landau dispersion relation (\ref{landau-dispersion-relation}).
The solutions in this case are shown in the left column of Fig.~\ref{u0}. Note that in the limit $k \to 0$, the higher modes are acoustic,
i.e., $\omega \propto k$, with the proportionality coefficients being complex numbers with
comparable real and imaginary parts \cite{derfler-sim-phys.fluids-1969}.

\begin{figure}
\includegraphics[width=11cm]{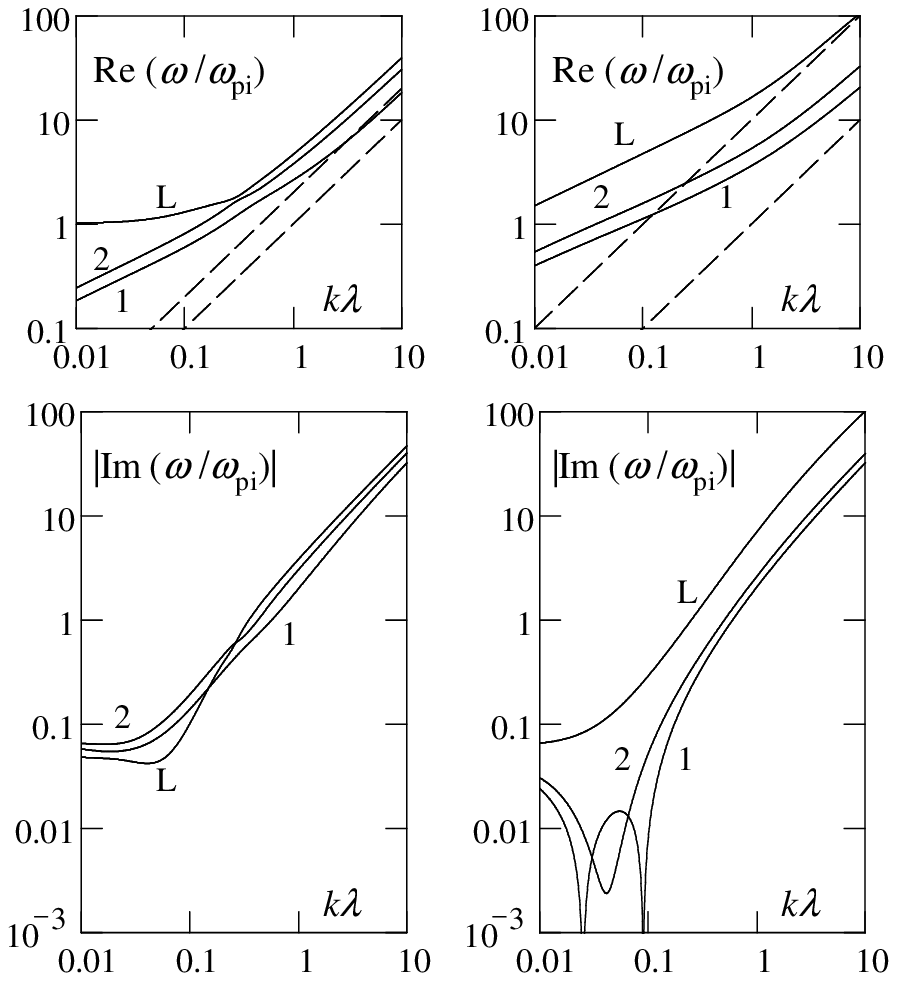}
\caption{Modes in the presence of flow.
The left 
and right columns
correspond to $u=2$ and $u=10$, respectively,
both are for $\zeta=0.1$ and downstream waves.
The notation of the modes is the same as in Fig.~\ref{u0}. 
The right column illustrates the instability 
of the first higher mode.
The upper and lower dash lines,
shown to guide the eye, correspond to
${\rm Re}(\omega)/k=v_{\rm f}$
and ${\rm Re}(\omega)/k=v_{\rm tn}$,
respectively.}
\label{u2u10}
\end{figure}

A finite $\zeta$ merely results in that ${\rm Im}(\omega)$ for any given mode
(including the ion Langmuir one) tends to a constant at $k \to 0$,
as shown in the right column of Fig.~\ref{u0}. This constant for all higher modes
is the same and equal to $-\nu$. For the ion Langmuir mode,
this constant differs by a factor of two and is equal to $-\nu/2$, for $\nu < 2 \omega_{\rm pi}$.

{\it Effect of field:} As a detailed discussion of our numerical results in light of the instability mechanism
is given in Sec.~\ref{interpretation}, in this section we only provide a brief description of what happens
in the presence of the field. 

Let us first consider the dispersion curves for downstream waves. 
The left column of Fig.~\ref{u2u10} shows that at $u=2$ and $\zeta=0.1$,
the ion Langmuir mode is no longer the least damped mode at large wave numbers,
as the first two higher modes have smaller decay rates.
(We determine which mode is the
ion Langmuir mode by looking at the behavior at $k \to 0$:
the ion Langmuir mode is the one that has a finite real part of
the frequency at $k \to 0$.)
The right column of Fig.~\ref{u2u10} illustrates that at $u=10$ and $\zeta=0.1$, the first higher mode is unstable
in a range of wave numbers, while the ion Langmuir mode remains stable at these parameter values.
It is not only the first higher mode that can become unstable, as we found numerically a large number of unstable
modes by increasing $u$ and decreasing $\zeta$ (in accordance with the analytical results of Secs.~\ref{proof}).

For upstream waves (for which the phase velocity vector is in the
direction opposite to the flow) we did not find any instability.
The general case of arbitrary angle of propagation can be mathematically reduced to the case of propagation
along or against the flow, as Eq.~(\ref{dispersion-relation}) contains ${\bf v}_{\rm f}$ 
only in the combination ${\bf k} \cdot {\bf v}_{\rm f}$.

\begin{figure}
\includegraphics[width=8.5cm]{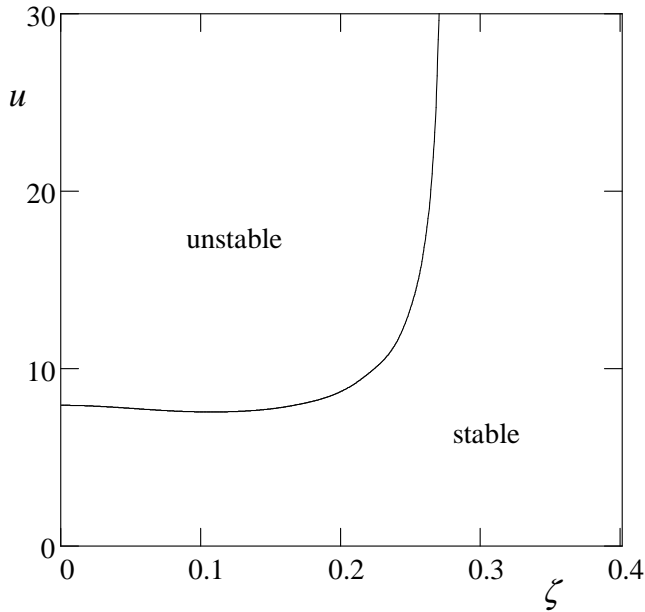}
\caption{Stability diagram. The instability region is bound within $u \gtrsim 8$ and $\zeta \lesssim 0.3$. 
Note that in the limit $\zeta \to 0$, $u = const$ within the instability region,
the growth rate of the instability tends to zero, so that a finite collision parameter is essential for the instability.}
\label{fig-stability-diagram}
\end{figure}

We calculated the instability region in the $(u, \zeta)$ space (see Fig.~\ref{fig-stability-diagram}).
One can see that the instability region is bound within $u \gtrsim 8$ and $\zeta \lesssim 0.3$. 
Here, two important comments need to be made. First, we found numerically that in the limit $\zeta \to 0$ and $u = const$ within the instability region,
the dimensionless growth rate tends to zero, ${\rm Im}(\omega/\omega_{\rm pi}) \to 0^{+}$.
Thus, a finite collision parameter is essential for the instability. Second, it is always downstream waves that
become unstable at the boundary of the instability region.

\subsection{Analytical proof of the instability existence}
\label{proof}
Let us analytically prove the existence of the instability
and show that the number of unstable modes can be made arbitrarily high by varying parameters $u$ and $\zeta$.
We consider Eq.~(\ref{dispersion-relation}) in the limit $v_{\rm tn} \to 0$, $\omega_{\rm pi} \to \infty$ at finite $v_{\rm f}$, $\nu$, $\omega$, $k$
and assume
that ${\bf k}$ is in the direction of the flow.
Then, the first term (unity) in Eq.~(\ref{dis-1}) is negligible
so that the dispersion relation takes the form $B=0$. Furthermore,
in Eq.~(\ref{dis-4}) the second term of the expression inside the square
brackets is negligible as well. 
In the resulting dispersion relation, 
let us consider the limit of large $k$. This allows us to neglect the unity
in the denominator of Eq.~(\ref{dis-3}) as well as the unity
in the first term of the right-hand side of Eq.~(\ref{dis-4})
and yields
\begin{subequations}
\label{analytic}
\begin{equation}
\omega=C \sqrt{k v_{\rm f} \nu},
\label{analytic-via-vf}
\end{equation}
where the numerical factor $C$ is given by
\begin{equation}
\int_0^\infty \exp \left( iC\xi - \frac{1}{2}i\xi^2 \right) \,d \xi =0.
\label{trans}
\end{equation}
\end{subequations}
Equation~(\ref{trans}) has an infinite number of solutions $C$, and they
all have positive real and imaginary parts. That is, we get {\it an infinite number
of unstable modes}. They correspond to downstream waves, 
as ${\rm Re}(\omega)>0$ for all solutions.
The most unstable solution, i.e., the one with
the largest ${\rm Im}(C)$, is
$C \approx 3.35+ 0.64i$.
The presence of an infinite number of unstable modes in the above limit means
that, by approaching this limit,
one can achieve an arbitrarily large number
of unstable modes.
This is consistent with our numerical analysis of
the dispersion relation (\ref{dispersion-relation}), as we numerically found a large number of unstable modes
at sufficiently large $u$ and sufficiently small $\zeta$ (as noted in Sec.~\ref{numerical-results}). 
Also, we have numerically verified
that the solutions described by Eq.~(\ref{analytic})
indeed satisfy Eq.~(\ref{dispersion-relation})
in the limit
considered above.
We also provide a derivation of Eq.~(\ref{analytic}) starting from our initial equations~(\ref{kinetic-equation})-(\ref{neutral-distribution})], 
see Sec.~\ref{interpretation}.

\subsection{Instability mechanism}
\label{interpretation}
First of all, let us emphasize that our instability is not a variation of any known collisionless instability (e.g., the bump-on-tail instability \cite{melrose-1986, nishikawa-wakatani-2000-p.92}),
as we show below that a collisionless one-component plasma with our steady-state velocity distribution (\ref{distribution}) is always stable.
We consider Eq.~(\ref{dispersion-relation}) in the limit of infinitely small ${\bf E}_0$
and $\nu$, but 
keep their ratio (which determines ${\bf v}_{\rm f}$ and hence the form of the steady-state velocity distribution) finite.
In this limit Eq.~(\ref{dispersion-relation})
simplifies to
\begin{equation}
1+\frac{1}{(k \lambda)^2} \int_0^\infty \frac{\xi \exp\left\{ i [\omega/(k v_{\rm tn})] \xi -\xi^2/2\right\}\, d{\xi}}{1+i\xi ({\bf k}\cdot {\bf v}_{\rm f})/(k v_{\rm tn})}=0.
\label{collisionless}
\end{equation}
This equation does not have unstable solutions, as we verified numerically. Note that this is in accordance with the
above stated fact
that in the limit $\zeta \to 0$, $u = const$
the instability growth rate tends to zero.

To explain the instability mechanism, let us start with the physics of the 
higher modes in the case of $u=0$ and $\zeta \to 0$ and then see how they can become unstable
in the presence of the field and collisions.

For $u=0$ and $\zeta \to 0$, the higher modes represent quasineutral oscillations at small wave numbers, as mentioned in the Introduction and explained 
in more detail below.
Let us first clarify that the term ``quasineutral'' here means that the neglect of the  $\bigtriangleup \phi$ term in 
Poisson's equation does not affect the higher modes, or, to put it differently,
that contributions from various velocity domains to the velocity integral determining the susceptibility
compensate for each other.
To demonstrate this, we first note that
the $\bigtriangleup \phi$ term in Poisson's equation contributes to Eq.~(\ref{landau-dispersion-relation}) by means of the first term
of Eq.~(\ref{landau-dispersion-relation}) (unity),
as seen from the derivation of Ref.~\cite{landau-j.phys.ussr-1946}.
This vacuum term, as seen directly from Eq.~(\ref{landau-dispersion-relation}), indeed does not affect the higher-order solutions $\omega$ in the limit $k \to 0$,
because of the $1/k^2$-factor in the second term of Eq.~(\ref{landau-dispersion-relation}) and the acoustic character of the higher modes ($\omega \propto k$) at $k \to 0$.
Appendix~\ref{quasineutrality-higher-modes} explicitly shows how contributions from various velocity domains to the velocity integral determining the susceptibility
compensate for each other.

These quasineutral waves are heavily Landau damped, as their phase velocity is of the order of the thermal velocity.
One way to eliminate the damping is, as noted in the Introduction, to apply an external driver
in order to create a population of trapped particles \cite{afeyan, johnston-tys-ghi-phys.plasmas-2009,
valentini-cal-per-phys.rev.lett-2011}. Another way, which is what is considered in this paper, 
is to induce field-driven flow. 

Let us now explain how growing quasineutral ion waves
become possible in the presence of field-driven ion flow, starting from our initial equations~(\ref{kinetic-equation})-(\ref{neutral-distribution}).

We consider the case where
the flow velocity is much larger than the thermal velocity of neutrals
and focus on the kinetics of ions with velocities in the range
$v_{\rm tn} \ll v_z \ll v_{\rm f}$.
Mathematically, this is equivalent to considering the limit of cold neutrals 
and simplifying the resulting steady-state distribution (\ref{simple-distr}) by taking its low velocity part,
\begin{eqnarray}
f_{0}({\bf v}) = \frac{n_0}{v_{\rm f}}\delta(v_x)\delta(v_y), \qquad  (v_{\rm f}\gg) v_z (\gg v_{\rm tn}) > 0, \nonumber \\
f_{0}({\bf v}) = 0, \qquad v_z<0.
\label{step-function}
\end{eqnarray}
By using this expression and considering downstream waves with $f-f_0= f_{\rm a}\exp(-i \omega t + i k z)$, 
$\phi = \phi_{\rm a} \exp(-i \omega t + i k z)$ (where the subscript ``${\rm a}$'' stands for ``amplitude''), we get the following linearized
kinetic equation:
\begin{eqnarray}
-i \omega f_{\rm a} + i k v_z f_{\rm a} + \frac{e E_0}{m}\frac{\partial f_{\rm a}}{\partial v_z}
-\frac{ike\phi_{\rm a} }{m} \frac{n_0}{v_{\rm f}}\delta({\bf v}) \nonumber \\
= -\nu f_{\rm a} + \nu \delta({\bf v}) \int f_{\rm a}({\bf v}') \, d{\bf v}'.
\label{kinetic-simple1}
\end{eqnarray}
Note that for this consideration to be valid, an additional term $ [i k e \phi_{\rm a} n_0/(m v_{\rm f}^2)] \exp(-v_z/v_{\rm f}) \delta (v_x) \delta (v_y)$ (for $v_z>0$) should be negligible,
as the latter appears when Eq.~(\ref{simple-distr}) is used instead of Eq.~(\ref{step-function}). We will analyze this condition, as well as other restrictive
conditions imposed below, later in this section.

Let us consider the ballistic case, where the right-hand side of Eq.~(\ref{kinetic-simple1}) is negligible.
This implies
\begin{equation}
\nu \left| \int f_{\rm a} \, d{\bf v} \right| \ll \frac{k e |\phi_{\rm a}|n_0}{m v_{\rm f}}
\label{condition1}
\end{equation}
and
\begin{equation}
\nu \ll {\rm max}\left\{|\omega|, {k v_z}\right\},
\label{condition2}
\end{equation}
which should be satisfied for the characteristic velocity $v_z$ of the solution $f_{\rm a}$; 
Eq.~(\ref{condition1}) follows from comparison of the terms containing the delta-function [i.e., the fourth and sixths terms in Eq.~(\ref{kinetic-simple1})],
and Eq.~(\ref{condition2}) follows from comparison of the terms not containing the delta-function [i.e., the first, second, and fifth terms
in Eq.~(\ref{kinetic-simple1}); we did not insert the third term of Eq.~(\ref{kinetic-simple1}) into Eqs.~(\ref{condition1}) and (\ref{condition2}),
because this term is the only remaining term, so that its magnitude is determined by the terms already included in Eqs.~(\ref{condition1}) 
and (\ref{condition2})].
The solution of Eq.~(\ref{kinetic-simple1}) with zero right-hand side
is
\begin{eqnarray}
f_{\rm a}=\frac{i k \phi_{\rm a}}{E_0} \frac{n_0}{v_{\rm f}}\exp \left[ \frac{m}{eE_0} \left(i\omega v_z - \frac{i k v_z^2}{2}\right)\right] \nonumber \\
\times \delta(v_x)\delta(v_y), \quad v_z >0, \nonumber \\
f_{\rm a}=0, \quad v_z<0.
\label{eigenfunction}
\end{eqnarray}
By substituting this solution to Poisson's equation 
$k^2 \phi_{\rm a}=(e/\epsilon_0) \int f_{\rm a} \, d{\bf v}$,
we get the dispersion relation
\begin{equation}
1-\frac{ie n_0}{\epsilon_0 E_0 k v_{\rm f}} \int_0^{\infty} \exp \left[ \frac{m}{eE_0} \left(i\omega v_z - \frac{i k v_z^2}{2}\right)\right] \, d{v_z}=0,
\label{vacuum-term}
\end{equation}
where the main contribution to the integral should be from the aforementioned range $v_{\rm tn} \ll v_z \ll v_{\rm f}$ [see Eq.~(\ref{step-function})]
for the consideration to be valid.
Note that the vacuum term in Eq.~(\ref{vacuum-term}) is the first term (unity), as it comes from the $k^2 \phi_{\rm a}$-term in Poisson's equation.

Let us consider the case where neglecting the vacuum term in Eq.~(\ref{vacuum-term}) does not affect some of its
solutions $\omega$ (the condition for this is obtained below). 
Neglecting the vacuum term, we get 
\begin{equation}
\omega=C\sqrt{\frac{eE_0k}{m}},
\label{analytic-via-field}
\end{equation}
where $C$ is given by Eq.~(\ref{trans}). Thus, we obtained exactly Eq.~(\ref{analytic}) [Eqs.~(\ref{analytic-via-field}) and (\ref{analytic-via-vf}) are equivalent] 
and hence its infinite set of unstable solutions. 
The above consideration explains the instability mechanism, as we identified the elementary terms and processes essential for the instability 
and how they work together.

Let us now show how the restrictive conditions imposed above determine the instability region in the $(u, \zeta, k)$ space.
Namely, in the following, we show that for the most unstable mode the above restrictive conditions can be summarized as:
\begin{equation}
1 \ll \frac{k v_{\rm f}}{\nu} \ll {\rm min} \left\{ u^2, \, \zeta^{-4/3}  \right\}.
\label{range}
\end{equation}
One can immediately see from condition~(\ref{range}) that it can only be satisfied when
\begin{equation}
u \gg 1, \quad \zeta \ll 1,
\label{basic-instability-conditions}
\end{equation}
which explains why the instability occurs at large $u$ and small $\zeta$ (see Fig.~\ref{fig-stability-diagram}).
As condition (\ref{range}) is for {\it neglecting} the effects that are not essential for the instability mechanism,
the instability range can be estimated as
\begin{equation}
1 \lesssim \frac{k v_{\rm f}}{\nu} \lesssim {\rm min} \left\{ u^2, \, \zeta^{-4/3}  \right\}.
\label{range-estimate}
\end{equation}

To derive Eq.~(\ref{range}), let us start with the condition for neglecting the vacuum term.
We rewrite Eq.~(\ref{vacuum-term}) as
\begin{eqnarray}
1-i \left( \frac{\omega_{\rm pi}}{\nu} \right)^2 \left( \frac{\nu}{v_{\rm f} k} \right)^{3/2} \nonumber \\
\times \int_0^\infty \exp \left( i \frac{\omega}{\sqrt{eE_0 k/m}} \xi - i \frac{\xi^2}{2} \right) \, d\xi=0,
\end{eqnarray}
and note that for the most unstable mode
\begin{equation}
{\rm Re} (\omega) \sim {\rm Im} (\omega) \sim \sqrt{eE_0 k/m}.
\label{omega-estimate}
\end{equation}
It follows that the vacuum term does not affect the most unstable mode
when 
\begin{equation}
\left( \frac{\omega_{\rm pi}}{\nu} \right)^2 \left( \frac{\nu}{v_{\rm f} k} \right)^{3/2} \gg 1.
\label{quasineutrality-condition}
\end{equation}

Let us now consider the condition that the characteristic velocity 
$v_z$ providing the main contribution to the integral in Eq.~(\ref{vacuum-term})
satisfies $v_{\rm tn} \ll v_z \ll v_{\rm f}$.
By using Eqs.~(\ref{omega-estimate}) we obtain that this velocity is $v_z \sim \sqrt{eE_0/(mk)}$, which gives
\begin{equation}
1 \ll \frac{k v_{\rm f}}{\nu} \ll u^2.
\label{diapazon1}
\end{equation}

Concerning the remaining conditions --- Eqs.~(\ref{condition1}) and (\ref{condition2}) and the condition of smallness of the term mentioned
just after Eq.~(\ref{kinetic-simple1}) --- they are automatically satisfied 
for the most unstable mode when Eqs.~(\ref{quasineutrality-condition}) and (\ref{diapazon1}) hold,
as can be easily shown. 

By combining Eqs.~(\ref{quasineutrality-condition})-(\ref{diapazon1})
we get Eq.~(\ref{range}). Note that another derivation of Eq.~(\ref{range}) [from Eq.~(\ref{dispersion-relation})]
is provided in Appendix~\ref{estimates}.

Let us now discuss, in light of the above findings, our numerical results shown in the right column of Fig.~\ref{u2u10}.
To do so, we first rewrite Eqs.~(\ref{analytic-via-field}) and (\ref{range-estimate}) in the dimensionless units used to plot Fig.~\ref{u2u10}:
\begin{equation}
\label{omega-dimensionless}
\frac{\omega}{\omega_{\rm pi}}=C \sqrt{u \zeta k \lambda},
\end{equation}
\begin{equation}
\frac{\zeta}{u} \lesssim k \lambda \lesssim {\rm min} \left\{ \zeta u, \frac{1}{u \zeta^{1/3}} \right\},
\label{range-dimensionless}
\end{equation}
where, to remind, $C \approx 3.35 + 0.64i$ for the most unstable mode.
One immediately sees that the real part of the frequency shown in the right column of Fig.~\ref{u2u10}
is in excellent agreement with Eq.~(\ref{omega-dimensionless}) (inside the instability range),
as the difference is within $15$\%. Remarkably, the square root dependence 
of ${\rm Re}(\omega)$ on $k$ is easily noticeable on the graph.
The imaginary part, however, does not show good agreement, which is not surprising because
the parameter values corresponding to the right column of Fig.~\ref{u2u10} are close to the instability boundary shown in Fig.~\ref{fig-stability-diagram}
[Eq.~(\ref{analytic-via-field}) is derived assuming that all effects that are not essential to the instability mechanism are negligible].
To verify this interpretation of the discrepancy, we run a test comparing ${\rm Im}(\omega)$ given by Eq.~(\ref{omega-dimensionless}) with ${\rm Im}(\omega)$ given by Eq.~(\ref{dispersion-relation})
for parameters $u$, $\zeta$ and $k \lambda$ 
very well satisfying inequalities (\ref{range-dimensionless})
and found excellent agreement.
Finally, comparison of Eq.~(\ref{range-dimensionless}) with the right column of Fig.~\ref{u2u10}
shows reasonable 
agreement, as Eq.~(\ref{range-dimensionless}) for the parameters of the right column of Fig.~\ref{u2u10} becomes $0.01 \lesssim k \lambda \lesssim 0.2$.

Let us also briefly discuss the wave number corresponding
to the maximum growth rate as well as the phase velocity of the most unstable mode.
Concerning the former, Eq.~(\ref{analytic-via-field})
shows that the growth rate increases with $k$, so the wave number corresponding
to the maximum growth rate can be estimated as the upper end of the instability wave number range (\ref{range-estimate}).
As regards the phase velocity of the most unstable mode, 
we combine Eqs.~(\ref{analytic-via-field}) and (\ref{range-estimate})
and thus obtain that this phase velocity 
varies with $k$ in the range
\begin{equation}
{\rm max}\left\{v_{\rm tn}, v_{\rm f}\zeta^{2/3}\right\} \lesssim \frac{{\rm Re}(\omega)}{k} \lesssim v_{\rm f}.
\end{equation}

Finally, let us emphasize that the unstable solutions (\ref{analytic-via-field}) are quasineutral. This already follows from the derivation itself,
as these solutions are obtained for the case where they are not affected by the neglect of the vacuum term (see above).
Let us, however, explicitly illustrate how high- and low-velocity particle contributions to the susceptibility compensate for each other.
The susceptibility is the second term in Eq.~(\ref{vacuum-term}), and we consider it as the sum of two parts,
one being the contribution from $v_z<{\rm Re}(\omega)/k$ and one from $v_z>{\rm Re}(\omega)/k$. By using the most
unstable solution $\omega \approx (3.35 + 0.64i) \sqrt{eE_0 k/m}$,
we obtain that the first part normalized by $(\omega_{\rm pi}/\nu)^2[\nu/(v_{\rm f} k)]^{3/2}$ is $\approx 0.1+0.04i$
and that the second part normalized as above is $\approx -0.1-0.04i$, so that they add up to zero.

\subsection{Constant mean free path case}
\label{role-of-the-form-of-the-collision-term}
It is already clear from Sec.~\ref{interpretation} that the instability mechanism is generic in the sense that it is not sensitive to a particular velocity
dependence for charge transfer collisions. Let us, however, explicitly demonstrate that the instability remains in the
constant mean free path case.
In the limit of cold neutrals the operator (\ref{cmfp}) simplifies to
\cite{kompaneets-kon-ivl-phys.plasmas-2007}
\begin{equation}
{\rm St}[f]=-\frac{vf}{\ell}+ \frac{\delta ({\bf v})}{\ell} \int f({\bf r}, {\bf v}') v' \, d{\bf v}'.
\label{cmfp-simplified}
\end{equation}
Then the steady-state distribution is
\begin{eqnarray}
f_0({\bf v}) = \frac{2n_0}{\pi v_{{\rm f}, \ell}} \exp \left( -\frac{v_z^2}{\pi v_{{\rm f}, \ell}^2} \right)\delta(v_x) \delta(v_y), \quad v_z>0, \nonumber \\
f_0 ({\bf v})= 0, \quad v_z<0,
\label{distr2}
\end{eqnarray}
where $v_{{\rm f}, \ell}= |\int {\bf v} f_0 \, d{\bf v}|/n_0 =\sqrt{2eE_0\ell /(\pi m)}$ is the flow 
velocity in the constant mean free path case.
The subsequent steps are exactly the same as in Sec.~\ref{interpretation},
and we come to the conclusion that the relation~(\ref{analytic-via-field})
exactly applies to the constant mean free path case as well (in the limit considered). 

\subsection{Role of the electron response}
\label{electron-response}

\begin{figure}
\includegraphics[width=9cm]{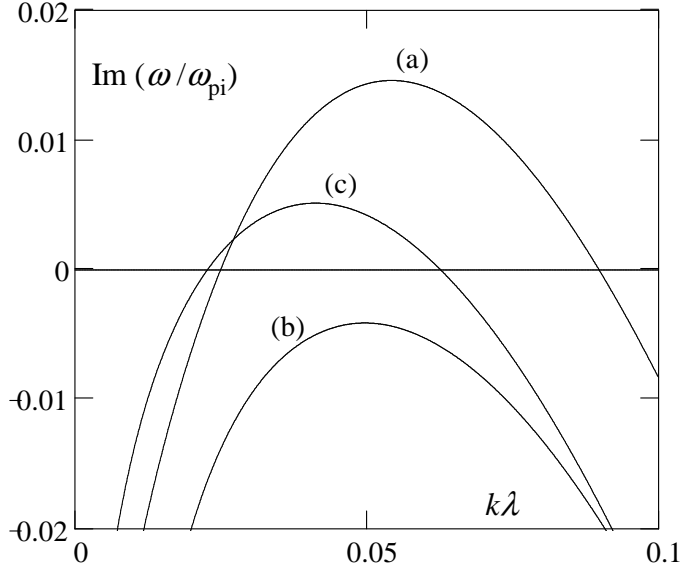}
\caption{Effect of the electron response.
(a) No response case: The unstable mode for $u=10$, $\zeta=0.1$ (i.e., for the same $u$ and $\zeta$ as in the right column of Fig.~\ref{u2u10}).
(b) The same mode for $T_{\rm e}/T_{\rm n} = 200$, with $u$ and $\zeta$ as above. 
(c) The same mode for $u=14$, $\zeta=0.05$ and $T_{\rm e}/T_{\rm n}$ as above.}
\label{fig-electron-response}
\end{figure}

Let us numerically show that the electron response for typical values of $T_{\rm e}$
does not shift the instability thresholds to unrealistic values. (How the electron response is included
is explained in Sec.~\ref{analysis}.) We choose a typical value $T_{\rm e}/T_{\rm n}=200$
corresponding to $k_{\rm B}T_{\rm e}= 5$~eV and $T_{\rm n}=300$~K. Performing calculations for $u=10$ and $\zeta=0.1$
(i.e., for the same values of $u$ and $\zeta$ as in the right column of Fig.~\ref{u2u10}), we find that the instability disappears 
(see Fig.~\ref{fig-electron-response}),
but the instability does not disappear
at, for instance, $u=14$ and $\zeta=0.05$ (see Fig.~\ref{fig-electron-response}). These values of $u$ and $\zeta$ are quite realistic, 
as will be seen in light of the discussion of Sec.~\ref{presheaths}.

We did not find
any other unstable mode in cases (a), (b) and (c) of Fig.~\ref{fig-electron-response}.
It should also be noted that the classical expression for the ion-acoustic instability~\cite{lifshitz-pitaevskii-1981}
requires a finite electron-to-ion mass ratio for the instability to occur,
while in our approach the electron mass is effectively zero.

\section{Discussion}
\label{discussion}

\subsection{Applicability limits}
\label{model-applicability-limits}
The applicability of our model is limited by the following factors: 
\begin{itemize}
\item[(i)] electron response to ion oscillations (addressed in Sec.~\ref{electron-response}),
\item[(ii)] inhomogeneity of the Boltzmann electron density
profile in the field ${\bf E}_0$ (discussed below),
\item[(iii)] assumption of a velocity-independent collision frequency for charge transfer collisions 
(addressed in Sec.~\ref{role-of-the-form-of-the-collision-term}), and
\item[(iv)] assumption that charge transfer is the dominant mechanism of ion scattering on neutrals (discussed below).
\end{itemize}
Concerning factor (iv), for argon at room temperature,
charge transfer indeed dominates when ion velocities exceed the thermal velocity of neutrals by a factor of $\sim 3$ or larger~\cite{kompaneets-phd}.
Thus, for argon at room temperature, factor (iv) should not substantially influence our instability, as the latter occurs at $u \gtrsim 8$.
As regards factor (ii), the corresponding inhomogeneity 
length is $L_{\rm e} = k_{\rm B}T_{\rm e}/(eE_0)$.
Our model can be applied when this distance is larger than both 
the ion-neutral collision length which is $v_{\rm f}/\nu$ (for $v_{\rm f} \gtrsim v_{\rm tn}$)
and the wavelength $2\pi/k$.
This imposes the following requirement:
\begin{equation}
\frac{T_{\rm e}}{T_{\rm n}} \gg {\rm max} \left\{ u^2, \, \frac{u \nu}{v_{\rm tn} k} \right\}.
\label{limitation1}
\end{equation}

Let us make estimates to find when the instability occurs in the case of a finite electron temperature.
To do so, we need to find when the instability mechanism explained in Sec.~\ref{interpretation}
is not substantially influenced by other effects, including the electron effects,
in a range of wave numbers. This means that
\begin{itemize}
\item condition (\ref{basic-instability-conditions}) is satisfied and
\item in at least a part of the wave number range (\ref{range})
both condition~(\ref{limitation1}) is met and
the effect of the Boltzmann response term $1/(k \lambda_{\rm e})^2$ on the most unstable mode is negligible.
\end{itemize}
As shown in Appendix~\ref{estimates}, conditions (a) and (b) above can be written as
\begin{equation}
u \gg 1, \quad \zeta \ll 1, \quad \frac{T_{\rm e}}{T_{\rm n}} \gg u^2,
\label{instability-restriction}
\end{equation} 
[note that the last inequality in Eq.~(\ref{instability-restriction})
coincides with one of the two conditions imposed by Eq.~(\ref{limitation1})].
Here two comments need to be made. 
First, the fact that the last inequality of Eq.~(\ref{instability-restriction}) is better satisfied for the stable curve (b)
than for the unstable curve (c) of Fig.~\ref{fig-electron-response} does not mean a contradiction between Eq.~(\ref{instability-restriction}) and Fig.~\ref{fig-electron-response},
as the first two inequalities of Eq.~(\ref{instability-restriction}) are better satisfied for the curve (c) than for the curve (b).
Second, estimates~(\ref{instability-restriction}) are for {\it neglecting} the effects that are not essential to the instability mechanism,
so the instability threshold on the electron temperature can be estimated from Eq.~(\ref{instability-restriction}) as 
\begin{equation}
\frac{T_{\rm e}}{T_{\rm n}} \sim u^2
\label{electron-threshold}
\end{equation} 
(for a given $u \gg 1$ and a given $\zeta \ll 1$). Thus, as the instability thresholds on $u$
and $\zeta$ in the absence of the electron effects 
are about 8 and 0.3, respectively (as given by the exact calculation presented in Fig.~\ref{fig-stability-diagram}),
we can summarize the instability conditions as
\begin{equation}
8 \lesssim u \lesssim \sqrt{\frac{T_{\rm e}}{T_{\rm n}}}, \quad \zeta \lesssim 0.3;
\label{instability-conditions-summary}
\end{equation}
Eq.~(\ref{instability-conditions-summary}) is not very accurate
(as seen from Fig.~\ref{fig-electron-response}) because 
the condition on $T_{\rm e}$ in Eq.~(\ref{instability-conditions-summary}) is merely an estimate
and not an exact result.

\subsection{Presheaths}
\label{presheaths}

Based on the above, we expect the instability to naturally occur in presheaths at sufficiently small, but still quite common, pressures,
as explained in the following. Let us see how the instability conditions~(\ref{instability-conditions-summary}) can be met in presheaths.

We first analyze the condition $\zeta \lesssim 0.3$.
By replacing $\nu$ by $v_{\rm f} \sigma n_{\rm n}$, where $ \sigma$ is the ion-neutral
cross section and $n_{\rm n}$ is the neutral number density, we can rewrite the condition $\zeta \lesssim 0.3$ as
\begin{equation}
P \lesssim \frac{\sqrt{k_{\rm B} T_{\rm n}}}{30 \sigma}\sqrt{\frac{n_0 e^2}{\epsilon_0}},
\label{pressure-1}
\end{equation}
where $P$ is the gas pressure (here we took the condition $u \gtrsim 8$ into account, i.e., we substituted $v_{\rm f} \sim 8 v_{\rm tn}$, because 
at larger $v_{\rm f}$ the restriction on $P$ is stronger). 
By substituting $T_{\rm n}=300$~K and 
$\sigma=6.5 \times 10^{-15}$~cm$^{2}$ (this value of $\sigma$ is 
derived in Ref.~\cite{kompaneets-kon-ivl-phys.plasmas-2007} for argon from the data of Ref.~\cite{hornbeck-phys.rev-1951}),
we get
\begin{equation}
P \lesssim (2\, {\rm Pa})\times \sqrt{\frac{n_0}{10^{14}\, {\rm m}^{-3}}}.
\label{pressure-2}
\end{equation}
The obtained condition can be easily satisfied in gas discharges, as there have been many experiments with pressures below 2~Pa and plasma densities
about or greater than $10^{14}$~${\rm m}^{-3}$ \cite{konopka-phd,
couedel-nos-zhd-phys.rev.lett-2009, nosenko-ivl-zhd-phys.plasmas-2009, pustylnik-ivl-tho-phys.plasmas-2009, 
nakamura-ish-phys.plasmas-2009}.
(Note that here $n_0$ denotes the local density in the presheath and not the density in the bulk of the discharge,
but for a rough estimate one can use the latter; see measurements on the 
presheath structure~\cite{hershkowitz-ko-wan-ieee.trans.plasma.sci-2005, 
oksuz-her-plasma-plasma.sources.sci.tech-2005}.)

Let us now discuss the condition $8 \lesssim u \lesssim \sqrt{T_{\rm e}/T_{\rm n}}$ in light of the Bohm criterion~\cite{riemann-j.phys.d.appl.phys-1991}.
First of all, we note that the Bohm criterion applies when the ion-neutral collision length is larger than the electron Debye length \cite{riemann-j.phys.d.appl.phys-2003}. 
This condition can be written as
\begin{equation}
P \lesssim \frac{1}{\sigma} \sqrt{\frac{n_0 e^2 k_{\rm B}T_{\rm n}^2}{\epsilon_0 T_{\rm e}}}.
\end{equation}
This is a weaker condition than Eq.~(\ref{pressure-1}) because they differ by the factor 
$30\sqrt{T_{\rm n}/T_{\rm e}}$, which is typically larger than unity, so the Bohm criterion applies when
Eq.~(\ref{pressure-1}) is satisfied.
According to the Bohm criterion,
the ion flow velocity reaches at least the Bohm speed $\sqrt{k_{\rm B} T_{\rm e}/m}$
at the sheath-presheath edge \cite{riemann-j.phys.d.appl.phys-1991}.
Then, the condition $8 \lesssim u \lesssim \sqrt{T_{\rm e}/T_{\rm n}}$ is met in 
a certain space region within the presheath
if 
\begin{equation}
\sqrt{\frac{T_{\rm e}}{T_{\rm n}}} \gtrsim 8.
\label{te}
\end{equation}
For $T_{\rm n}=300$~K, this means $k_{\rm B}T_{\rm e} \gtrsim 1.6$~eV, which is often satisfied.

The instability may have significant implications, as
presheaths and sheaths are important to plasma physics and technology \cite{lieberman-1994, hershkowitz-phys.plasmas-2005}. 
First, the instability may result in a flow turbulence or various dynamic structures and thus lead to the appearance of strong electric fields varying on the ion time scale. 
An alternative is the formation of a static structure that suppresses the instability.
In an extreme scenario, the instability may significantly affect the whole discharge or even switch it off.
The above effects cannot be modeled using the hydrodynamic approach since the latter ignores the higher modes. 
(Inaccuracy of hydrodynamic modeling of plasma boundary layers is illustrated in, e.g., Ref.~\cite{kuhn-rie-jel-phys.plasmas-2006}.)

\subsection{Presheath instability experiments}
\label{experiments}
Instabilities in presheaths have been observed in the presence of 
two ion species \cite{hershkowitz-ko-wan-ieee.trans.plasma.sci-2005, hershkowitz-yip-sev-phys.plasmas-2011}.
The ion-kinetic instability described in this paper, in contrast, can occur even for a single ion species plasma.  
Interestingly, measurements of Ref.~\cite{hershkowitz-yip-sev-phys.plasmas-2011}
indicate that the presheath instability did not disappear
when only one ion species was present (although in this case the instability amplitude was reduced; see Fig.~10 of Ref.~\cite{hershkowitz-yip-sev-phys.plasmas-2011}).
To explain this, the presence of unidentified impurity ions was suggested \cite{hershkowitz-yip-sev-phys.plasmas-2011}.
We suggest that these results might also be explained by
the instability of the higher modes.

Let us explicitly suggest experimental conditions to observe the instability described in this paper.
Concerning the choice of gas, argon is an excellent candidate, as in this gas at room temperature,
charge transfer is indeed the dominant
mechanism of ion scattering when ion velocities exceed the thermal velocity of neutrals by a factor of $\sim 3$ or larger~\cite{kompaneets-phd}.
In addition, an argon plasma well represents a single ion species plasma,
as the impurity degree due to ions other than Ar$^+$ (e.g., Ar$^{++}$ and Ar$_2^+$) is usually quite small \cite{olthoff-bru-rad-j.app.phys-1994.pdf}.
Concerning the gas pressure, it should satisfy Eq.~(\ref{pressure-2}). An example of an experiment well satisfying this condition
is Ref.~\cite{lee-her-sev-appl.phys.lett-2007}. [Note, however, that at extremely small pressures, the instability may not be detectable,
as its growth rate tends to zero at $\zeta \to 0$ (Secs.~\ref{numerical-results} and \ref{interpretation}).]
It may be also necessary to take measures to increase the electron temperature,
as our estimate of the instability threshold on the electron temperature yields
$k_{\rm B}T_{\rm e} \sim 1.6$~eV [Eq.~(\ref{te})] so that the actual threshold
may differ by a factor of a few.
These measures can be: (i) decreasing the gas pressure \cite{konopka-phd}, (ii) decreasing the rf peak-to-peak voltage \cite{konopka-phd},
and (iii) using a Maxwell demon \cite{mackenzie-tay-coh-appl.phys.lett-1971, hershkowitz-yip-sev-phys.plasmas-2011}.

\subsection{Dusty plasmas}

Another implication of the instability described in this paper is that 
this instability may affect the interparticle interaction \cite{kompaneets-kon-ivl-phys.plasmas-2007,
kompaneets-mor-ivl-phys.plasmas-2009} and ion drag force \cite{ivlev-zhd-khr-phys.rev.e-2005} in dusty plasmas \cite{vladimirov-ost-phys.rep-2004, morfill-ivl-rev.mod.phys-2009, ishihara-j.phys.d.appl.phys-2007, shukla-eli-rev.mod.phys-2009}.
In particular, our results imply that
the expression for the shielding of a dust particle in the presence
of ion flow given by Eq.~(6) of Ref.~\cite{kompaneets-kon-ivl-phys.plasmas-2007}
is only valid when the ratio of the ``field-induced Debye length'' (defined in Ref.~\cite{kompaneets-kon-ivl-phys.plasmas-2007}) 
to the collision length is larger than a certain threshold, which is 
supposedly close to 0.3, i.e., to that in the BGK case
(otherwise, the linear response formalism does not apply because of the instability of the steady state).
The results of Ref.~\cite{kompaneets-mor-ivl-phys.plasmas-2009} are unaffected because, in that work, the subthermal flow regime was considered. 
The resulting change in the interaction between dust particles may affect 
their self-organization and dynamics \cite{totsuji-kis-tot-phys.rev.lett-2002,
lampe-joy-gan-ieee.trans.plasma.sci-2005,
ashwin-gan-phys.rev.e-2009, farokhi-sha-kou-phys.plasmas-2009, koukouloyannis-kou-phys.rev.e-2009, upadhyaya-hou-mis-phys.lett.a-2010, qiao-mat-hyd-ieee.trans.plasma-sci-2010, 
carstensen-gre-pie-phys.plasmas-2010, bonitz-hen-blo-rept.prog.phys-2010,
sheridan-wel-phys.rev.e-2010, vaulina-kos-khr-phys.rev.e-2010, liu-i-phys.rev.e-2010}.

\section{Conclusion}
We found a remarkable type of instability, which can be triggered in 
a weakly ionized plasma in the presence of ion flow driven by a dc electric field. 
We showed that the dc field and ion-neutral collisions can substantially
modify the physics of the ion higher-order Landau modes so that some of them can 
become unstable.
The instability is of broad relevance to gas discharge physics, as dc fields are common in gas discharges.
In particular, the instability is expected to naturally occur in presheaths under quite common conditions [Eqs.~(\ref{pressure-2}) and (\ref{te})]
and thus affect sheaths and discharge structures.
In even broader context, our study shows that the often disregarded higher-order 
Landau modes can in fact play a crucial role in the presence of an electric field.

\begin{acknowledgments}
The authors thank the Editorial Board Member for a thorough review and helpful comments.
R.K. acknowledges
the support of the 
Science
Foundation for Physics within the University of Sydney
and the hospitality of the Max-Planck-Institut f\"ur extraterrestrische Physik,
Germany. A.V.I. and G.E.M. received funding from the European Research Council under the European Union's Seventh
Framework Programme (FP7/2007-2013) / ERC Grant agreement 267499.
The work was partially supported by the Australian Research Council. 

\end{acknowledgments}

\appendix

\section{Dimensionless form of Eq.~(\ref{dispersion-relation})}
\label{dispersion-relation-dimensionless}
The dispersion relation (\ref{dispersion-relation}) can be written in our dimensionless units for a numerical analysis as
\begin{eqnarray}
1+\frac{1}{(k\lambda)^2} \left( 
1-\frac{\zeta}{k\lambda}
\int_0^{\infty} \exp(-H) \,d\xi
\right)^{-1}
\nonumber \\
\times \int_0^\infty
\frac{\xi\exp(-H)\, d\xi}{1+iu\xi\cos\theta}=0,
\nonumber \\
H=\frac{1}{k\lambda}
\left( 
\zeta \xi -\frac{i\omega \xi}{\omega_{\rm pi}}
+\frac{1}{2}iu\zeta \xi^2 \cos\theta
\right)
+\frac{\xi^2}{2},
\label{dispersion-relation-dimensionless-equation}
\end{eqnarray}
where $\theta$ is the angle between ${\bf k}$ and ${\bf E}_0$.

\section{Quasineutrality of the higher modes}
\label{quasineutrality-higher-modes}
This Appendix illustrates, for $u=0$ and $\zeta \to 0$, the quasineutral character of the higher modes at small wave numbers.
Namely, we explicitly show how contributions from various velocity domains to the velocity integral determining the susceptibility
compensate for each other.
The susceptibility, i.e., the second term in Eq.~(\ref{landau-dispersion-relation}), can be written as \cite{landau-j.phys.ussr-1946}:
\begin{equation}
\chi =-\frac{e^2}{m \epsilon_0 k^2} \int \limits_U \frac{df_{0,z}(v_z)}{dv_z} \frac{dv_z}{v_z - \omega/k},
\label{susceptibility}
\end{equation}
where the wave number is in the direction of the $z$ axis, $f_{0,z}(v_z)= (2 \pi)^{-1/2} (n_0/v_{\rm tn}) \exp [ -v_z^2/(2v_{\rm tn}^2)]$
is the velocity distribution integrated over $v_x$ and $v_y$,
and $U$ is any contour in the complex $v_z$ plane
passing from $v_z = -\infty$ to $+\infty$ below the singular point $v_z=\omega/k$.
This integral can be written as the sum of the integral on the real axis 
and the contribution from the pole of the integrand [for ${\rm Im}(\omega)<0$]. Dividing the integral on the real axis into two parts,
one over $v_z$ smaller than the phase velocity ${\rm Re}(\omega)/k$ and one over the remaining interval,
we can write
\begin{equation}
\chi=\frac{1}{(k \lambda)^2}(J_1 + J_2 + J_3),
\label{j1j2j3}
\end{equation}
where
\begin{eqnarray}
J_1=-\frac{v_{\rm tn}^2}{n_0} \int_{-\infty}^{{\rm Re}(\omega)/k}  \frac{df_{0,z}(v_z)}{dv_z} \frac{dv_z}{v_z - \omega/k}, \\
J_2=-\frac{v_{\rm tn}^2}{n_0} \int_{{\rm Re}(\omega)/k}^{\infty}  \frac{df_{0,z}(v_z)}{dv_z} \frac{dv_z}{v_z - \omega/k}, \\
J_3=-\frac{2\pi i v_{\rm tn}^2}{n_0} \left. \frac{df_{0,z}(v_z)}{dv_z} \right|_{v_z=\omega/k},
\end{eqnarray}
the integration in the expressions for $J_1$ and $J_2$ is performed on the real axis and $J_3$ is the contribution from the pole.
By substituting here the first higher mode solution
$\omega/(kv_{\rm tn}) \approx 3.60-1.73i$ (at $k \to 0$),
we get $J_1 \approx -0.032-0.06i$, $J_2 \approx 5 \times 10^{-5}-3\times 10^{-4}i$, 
$J_3 \approx 0.032+0.06i$, so that it is explicitly seen that they add up to zero and that, as a side note,
$J_2$ is almost unimportant here. For the second mode, we have $\omega/(k v_{\rm tn}) \approx 4.47-2.87i$,
$J_1 \approx -0.012-0.035i$, $J_2 \approx 5 \times 10^{-7}-6\times 10^{-6}i$, 
$J_3 \approx 0.012+0.035i$.

\section{Estimates}
\label{estimates}
This appendix provides estimates for the instability region
and thus derives Eqs.~(\ref{range}) and (\ref{instability-restriction}).
We first consider the ``pure'' case where all terms
that are not essential for the instability are neglected.
This allows us to obtain estimates for the magnitudes of the essential terms.
We then compare them with the neglected terms to obtain conditions under which the neglected
terms are indeed negligible.

The ``pure'' case can be considered by omitting all terms in Eq.~(\ref{dispersion-relation}) that were neglected in Sec.~\ref{proof}. These terms are: 
(i) the unity in round brackets in Eq.~(\ref{dis-4}),
(ii) the second term of the
expression inside the square brackets in Eq.~(\ref{dis-4}),
(iii) the unity in the denominator in Eq.~(\ref{dis-3}), and
(iv) the first term (unity) in Eq.~(\ref{dis-1}).
Therefore, the main contribution to the integral in Eq.~(\ref{dis-3}) is from 
\begin{equation}
\eta \sim \frac{\nu}{|\omega|} \sim \sqrt{\frac{\nu}{k v_{\rm f}}}.
\label{eta-estimate}
\end{equation}
To derive this, we expressed $\omega$ via $k$ using Eq.~(\ref{analytic}); we assume ${\rm Re} (C) \sim {\rm Im}(C) \sim 1$,
as we consider the most unstable mode; we also assume that ${\bf k}$ is in the direction of ${\bf E}_0$.
The above estimate of $\eta$ yields 
\begin{equation}
|B| \sim \left( \frac{\nu}{k v_{\rm f}}\right)^{3/2}.
\end{equation} 
This result is obtained by replacing $\eta$ and $d{\eta}$ by the estimate (\ref{eta-estimate}) and substituting unity for the exponent. 
Analogously, for the second term in Eq.~(\ref{dis-1}) we get
\begin{equation}
\frac{\omega_{\rm pi}^2}{\nu^2} \left|\frac{B}{1-A}\right|\sim \frac{\omega_{\rm pi}^2}
{\sqrt{\nu k^3 v_{\rm f}^3}} {\rm min}\left\{ 1, \, \sqrt{\frac{kv_{\rm f}}{\nu}}\right\}.
\label{second-term-estimate}
\end{equation}

Let us now make a comparison with the magnitudes of the non-essential terms (i)-(iv).
The term (i) is negligible when $|\omega| \gg \nu$. This is equivalent to $k \gg \nu/v_{\rm f}$.
The term (ii) can be omitted when $k \ll v_{\rm f} \nu/v_{\rm tn}^2$.
The term (iii) does not play any role when it is smaller than the other term 
in the denominator of Eq.~(\ref{dis-3}) with $\eta$ replaced by estimate (\ref{eta-estimate}). 
This gives $k \gg \nu/v_{\rm f}$, which coincides with the condition
for the neglect of the term (i).
Finally, the term (iv) can be neglected when it is smaller than the right-hand side of Eq.~(\ref{second-term-estimate}). 
The right-hand side of Eq.~(\ref{second-term-estimate}) can be simplified using the condition $k \gg \nu/v_{\rm f}$ for the neglect of the term (i). The result is that the term (iv) is negligible when 
$k \ll (\nu/v_{\rm f})(\omega_{\rm pi}/\nu)^{4/3}$.
By combining the above conditions, we get Eq.~(\ref{range}).

Analogously, we obtain that the electron response term $1/(k \lambda_{\rm e})^2$, added to the left-hand side of Eq.~(\ref{dis-1}), is unimportant when:
\begin{equation}
k \gg \frac{v_{\rm f}^3 \nu m^2}{(k_{\rm B}T_{\rm e})^2}.
\label{range-electron-response}
\end{equation}

Let us now see when conditions (\ref{range-electron-response}) and (\ref{limitation1})
are met in at least a part of the wave number range (\ref{range}) assuming that condition (\ref{basic-instability-conditions}) is satisfied.
As conditions (\ref{range-electron-response}) and (\ref{limitation1}) impose lower (and not upper) limits on the wave number,
we only need to find when conditions (\ref{range-electron-response}) and (\ref{limitation1}) are met
at the upper end of the wave number range~(\ref{range}).
Concerning Eq.~(\ref{range-electron-response}), it is met at the upper end of the wave number range~(\ref{range}) when
\begin{equation}
\frac{T_{\rm e}}{T_{\rm n}} \gg {\rm max} \left\{ 
u, u^2\zeta^{2/3} \right\}.
\label{compatibility-1}
\end{equation}
As regards Eq.~(\ref{limitation1}), it is met at the upper end of the wave number range~(\ref{range}) when $T_{\rm e}/T_{\rm n} \gg u^2$.
This inequality is a stronger condition than Eq.~(\ref{compatibility-1}).
Thus, we arrive at Eq.~(\ref{instability-restriction}).

\end{document}